\documentstyle[jkas]{article}

\beginpage{213}
\endpage{223}
\year{2010}\volume{43}\month{December}

\runningauthor {J.-S. HA ET AL.}
\runningtitle{EFFECT OF FIR FLUXES ON YSOS}

\month{December} \year{2010} \volume{43} \issueno{6}
\beginpage{213}\endpage{223}
\date{Received November 09, 2010; Revised December 02, 2010; Accepted December 02, 2010}

\begin{document}
\title{EFFECT OF FIR FLUXES ON CONSTRAINING PROPERTIES OF YSOS}
\author{Ji-Sung Ha$^{1}$, Jeong-Eun Lee$^{1}$, and Woong-Seob Jeong$^{2}$   }
\address{$^1$ Department of Astronomy and Space Science, Sejong University, Kunja-dong 98, Kwangjin-gu, Seoul 143-747, Korea\\
 {\it E-mail : uranoce27@sejong.ac.kr and jelee@sejong.ac.kr}}
\address{$^2$ Korea Astronomy and Space Science Institute,
61-1 Whaam-dong, Yuseong-Gu, Daejeon 305-348, Korea\\ {\it E-mail : jeongws@kasi.re.kr}}

\address{\normalsize{\it (Received November 09, 2010; Revised December 02, 2010; Accepted December 02, 2010)}}
\abstract{Young Stellar Objects (YSOs) in the early evolutionary stages are very embedded, and thus
they emit most of their energy at long wavelengths such as far-infrared (FIR) and submillimeter (Submm).
Therefore, the FIR observational data are
very important to classify the accurate evolutionary stages of these embedded YSOs, and to better constrain their physical parameters in the dust continuum modeling. We selected 28 YSOs,
which were detected in the AKARI Far-Infrared Surveyor (FIS), from the Spitzer c2d legacy YSO catalogs to test the effect of FIR fluxes on the classification of their evolutionary stages and on the
constraining of envelope properties, internal luminosity, and UV strength of the Interstellar Radiation Field (ISRF). According to our test, one can mis-classify the evolutionary stages of YSOs,
especially the very embedded ones if the FIR fluxes are not included. In addition, the total amount of heating of YSOs can be underestimated without the FIR observational data.}

\maketitle


\section{INTRODUCTION}

Stars are born in their parental molecular cloud cores through gravitational contraction. After the gravitational collapse begins, a protostar with its circumstellar disk forms at the center of the
collapsing core. Although the protostar/disk system is embedded in the dense infalling envelope at the early phase of evolution of star, the envelope gradually disappears as the protostar
evolves to the main-sequence star. The Spectral Energy Distribution (SED) of a Young Stellar Object (YSO) is contributed by the protostar at Near-Infrared wavelengths whose
temperature ranges from 2000 K to 4000 K, the disk at Mid-Infrared (MIR) wavelength which has a temperature gradient with radius, and the envelope at
Far-Infrared (FIR, 40 $<$ $\lambda$ $<$ 350 $\micron$) or Submillimeter (Submm, $\lambda$ $\geq$ 350 $\micron$) wavelengths.
Therefore, the analysis of the SED of a YSO can provide a way to identify the evolutionary stage of the YSO.
In the SED analysis, however, we need to understand the effect of the interstellar radiation field (ISRF) since the envelope is heated by the ISRF as well as the internal source luminosity.

There is a commonly known scheme classifying the evolutionary stages of YSOs based on the shape of the SED of a YSO (Lada 1987; Andr\'{e} et al. 1993).
In this proposed scheme, the Class 0 object is in the earliest stage after the
collapse begins. The Class 0 object already harbors a central heating source, but most of the mass is still in the dense envelope of gas and dust.
Therefore, all emission from the central source is reprocessed through the envelope, which is almost isothermal with T$\sim$30 K (Young et al. 2003).

Class I object is still heavily embedded, but emission from the central protostar and disk escapes through the outflow cavities.
As a result, the peak of the SED shows at MIR wavelengths.

As the envelope material accretes to the central star through the disk and/or is cleared out by the outflows,
the protostar/disk system becomes the dominant mass reservoir.
Therefore, Class II objects are less embedded and have the emission excesses at near and mid-infrared wavelengths by dusty disk.
The Class III objects emit energy mainly from central protostars and have only a small amount of gas in disks. The Class II
and the Class III are often classified as classical T Tauri Stars and weak-line T Tauri stars, respectively (White et al. 2007).

As mentioned, in the early embedded phases, the radiation from central source undergoes the absorption/re-emission processes by the heavy envelope. As a result, significant energy is emitted at FIR
wavelengths. Andr\'{e} et al. (2000) noted that deeply embedded Class 0, whose energy peak is located at beyond 100 $\micron$, were often missed by IRAS observations unlike the Class I objects,
which reveal the peak energy emission at the wavelength from 12 to 100 $\micron$. Therefore, the FIR data are crucial to study the evolutionary stages of embedded sources.

The envelope is also heated by the ISRF so that the SED at FIR and Submm is affected by the energy input from the ISRF as well as the central source.
Since the heating from the ISRF also affects the total luminosity of a YSO and the shape of the dust temperature profile,
it is very important to constrain the strength of the ISRF, and thus
the internal luminosity of the central heating source. The external heating by the ISRF compared to the internal heating is not negligible especially in the early embedded phase and in low luminosity sources (Dunham et al. 2008).

The AKARI/Far-Infrared Surveyor (FIS) All-Sky Survey Bright Source Catalog combined with precedented catalogs such as those by the Spitzer observations provides a useful tool
to investigate the importance of the FIR fluxes in the calculations of the evolutionary indicators of YSOs, such as $T_{bol}$ and $L_{bol}$/$L_{smm}$.
In this paper, AKARI/FIS All Sky Survey Bright Source Catalog is used to investigate the importance of the FIR fluxes on the evolutionary indicators of YSOs, $T_{bol}$ and $L_{bol}$/$L_{smm}$. In addition,
we model the SED of one YSO to study how the FIR fluxes can constrain the internal luminosity and the ISRF strength around the YSO.

This paper is structured as follows.
In section 2 and 3, the catalogs used in this study and the criteria adopted to select our sample YSOs are introduced. The evolutionary indicators of YSOs are summarized in section 4.
We calculate those evolutionary indicators of our samples and present the results in section 5. The dust continuum modeling of the SED of one sample YSO is presented in section 6,
and the summary and conclusions are followed in section 7.

\section{CATALOGS}

In this study, we use various infrared point source catalogs. All of them, except for one from the observations with Spitzer Space Telescope, were obtained through the all sky survey.

\subsection{InfraRed Astronomy Satellite (IRAS) Catalog}

IRAS was launched on January 26, 1983 with a cooled 57 cm telescope for all-sky survey in 4 bands
(12 $\micron$, 25 $\micron$, 60 $\micron$, and 100 $\micron$).
The typical aperture size in each band is 3.5\arcmin, 3.5\arcmin, 3.6\arcmin, and 4.7\arcmin, respectively (Wheelock et al. 1994).
Through IRAS all-sky survey, 72\% of the sky was observed with three or more hours-confirming scans,
and 95\% was observed with two or more hours-confirming scans (Neugebauer et al. 1984).

\subsection{Two Micron Sky Survey (2MASS) Catalog}

Observations of the Two Micron Sky Survey began in 1997 and were accomplished in 2001
with the two telescopes located in the northern and southern hemispheres.
The survey was performed for the entire sky using photometry system of 3 bands,
i.e., J (1.235 \micron), H (1.662 \micron), Ks (2.159 \micron).
The aperture size in all bands is 2.5\arcsec (Skrutskie et al. 2006).

\subsection{The c2d Spitzer legacy YSO Catalogs}

The Spitzer Space Telescope (SST) was launched in 2003 which has three instruments, i.e.,
the Infrared Array Camera (IRAC), the Infrared Spectrograph (IRS), and the Multiband Imaging Photometer for Spitzer (MIPS).
The IRAC has four bands (IRAC1, IRAC2, IRAC3, and IRAC4), which have central wavelengths
at 3.6 $\micron$, 4.5 $\micron$, 5.8 $\micron$, and 8.0 $\micron$, respectively,
and the beam FWHM is 1.66\arcsec, 1.72\arcsec, 1.88\arcsec, and 1.98\arcsec\ in each band.
Two bands of the MIPS, MIPS1 and MIPS2, have effective wavelengths at 24 $\micron$ and 70 $\micron$, respectively,
and the beam FWHM in each band is 6\arcsec and 18\arcsec.
The YSO catalogs used for this study were provided by the Spitzer Legacy Program
``From Molecular Cores to Planet Forming Disks''(c2d; Evans et al. 2003).
The c2d YSO catalogs provide the IRAC and MIPS fluxes of identified YSOs in five nearby, large molecular clouds (Serpens, Lupus, Chamaeleon II, Ophiucus, and Perseus)
and 58 small cores.

\subsection{AKARI (ASTRO-F)/FIS Bright Source Catalog}

AKARI, launched on February 2004, is a cooled 67 cm telescope with two focal plane instruments :
the Far Infrared Surveyor (FIS) and the Infrared Camera (IRC).
AKARI/FIS observed around 94$\%$ of the whole sky twice or more with a higher
spatial resolution and a wider wavelength coverage than those of the previous all-sky mission,
IRAS (Kawada et al. 2007).
Due to the higher resolution of AKARI/FIS observations, AKARI/FIS could resolve the background and the source more effectively than IRAS (Jeong et al. 2007).
The FIS performed the all-sky survey with 4 photometric bands (N60, WIDE-S, WIDE-L, and N160)
which have an effective wavelength at 65 $\micron$, 90 $\micron$, 140 $\micron$, and 160 $\micron$, respectively.
The beam FWHM of each band is 37$\arcsec$, 39$\arcsec$, 58$\arcsec$, and 61$\arcsec$, respectively (Kawada et al. 2007).

\begin{deluxetable}{crrccr}
\tablecolumns{6}
\tablewidth{0pt}
\small
\tablecaption{Properties of YSOs selected for this study}
\tablehead{
\colhead{Index} & \colhead{c2d Region\tablenotemark{1}} & \colhead{Spitzer Source Name\tablenotemark{2}} &  \colhead{R.A.(2000.0)}   & \colhead{Decl.(2000.0)}    & \colhead{AKARI Name}}
\startdata
1 & Ophiuchus & J163133.4-242737 &       247.891 &      -24.4594 & 1631337-242734 \\
2 & Ophiuchus & J163135.6-240129 &       247.899 &      -24.0247 & 1631357-240129 \\
3 & Lupus I & J153848.4-344038 &       234.703 &      -34.6759 & 1538486-344033 \\
4 & Perseus & J032538.8+304406 &       51.4102 &       30.7361 & 0325384+304410 \\
5 & Perseus & J032637.5+301528 &       51.6560 &       30.2584 & 0326374+301530 \\
6 & Perseus & J032834.5+310051 &       52.1426 &       31.0154 & 0328342+310055 \\
7 & Perseus & J032837.1+311331 &       52.1533 &       31.2267 & 0328368+311336 \\
8 & Perseus & J032845.3+310542 &       52.1875 &       31.0966 & 0328450+310548 \\
9 & Perseus & J032923.5+313330 &       52.3481 &       31.5575 & 0329235+313327 \\
10 & Perseus & J033121.0+304530 &       52.8372 &       30.7585 & 0331209+304531 \\
11 & Perseus & J033229.2+310241 &       53.1213 &       31.0448 & 0332291+310241 \\
12 & Perseus & J033312.8+312124 &       53.3031 &       31.3563 & 0333127+312123 \\
13 & Perseus & J033317.9+310932 &       53.3258 &       31.1597 & 0333182+310935 \\
14 & Perseus & J033327.3+310710 &       53.3642 &       31.1198 & 0333274+310711 \\
15 & Perseus & J034255.9+315842 &       55.7321 &       31.9785 & 0342557+315843 \\
16 & Serpens & J182815.3+000-243 &       277.063 &    -0.0444 & 1828152-000240 \\
17 & Serpens & J182858.1+001724 &       277.242 &      0.2896 & 1828580+001723 \\
18 & Serpens & J182906.2+003043 &       277.277 &      0.5116 & 1829064+003042 \\
19 & Serpens & J182932.0+011843 &       277.384 &       1.3119 & 1829322+011843 \\
20 & DC275.9+1.9 & J094645.9-510610 &       146.691 &      -51.1028 & 0946457-510610 \\
21 & DC302.1+7.4 & J124540.0-552522 &       191.417 &      -55.4212 & 1245402-552516 \\
22 & L100 & J171603.2-205657 &       259.013 &      -20.9489 & 1716030-205656 \\
23 & L1251 & J223031.9+751409 &       337.633 &       75.2351 & 2230318+751406 \\
24 & L723 & J191753.7+191220 &       289.474 &       19.2054 & 1917539+191219 \\
25 & Lupus III & J160918.1-390453 &       242.327 &      -39.0816 & 1609184-390454 \\
26 & Lupus IV & J160044.5-415531 &       240.185 &      -41.9242 & 1600443-415527 \\
27 & Lupus IV & J160115.5-415235 &       240.315 &      -41.8752 & 1601157-415231 \\
28 & Lupus IV & J160221.6-414054 &       240.589 &      -41.6811 & 1602213-414052

\label{tab1}
\enddata
\tablenotetext{1}{Name of the region where each source is located.}
\tablenotetext{2}{All source names are preceded by the prefix ``SSTc2d.''}
\end{deluxetable}

\begin{deluxetable}{cccccccccccc}
\tiny
\setlength{\tabcolsep}{1.25mm}
\tablecolumns{10}
\tablewidth{0pt}
\tablecaption{2MASS and Spitzer flux densities of YSOs selected for this study}
\tablehead{
\colhead{} & \colhead{J(1.25 \micron)} & \colhead{H(1.65 \micron)} & \colhead{$K_{s}$(2.17 \micron)} & \colhead{IRAC1(3.6 \micron)} & \colhead{IRAC2(4.5 \micron)} & \colhead{IRAC3(5.8 \micron)} & \colhead{IRAC4(8.0 \micron)} & \colhead{MIPS1(24 \micron)} & \colhead{MIPS2(70 \micron)} \\
\colhead{Index} & \colhead{mJy} & \colhead{mJy} &  \colhead{mJy}   & \colhead{mJy}    & \colhead{mJy} & \colhead{mJy}   & \colhead{mJy} & \colhead{mJy} & \colhead{mJy}}
\startdata

 1 &        323$\pm$6.84 &          515$\pm$27 &        602$\pm$13.3 &          620$\pm$34.8 &        540$\pm$29.7 &        489$\pm$23.7 &            673$\pm$33.9 &       1690$\pm$158 &       2560$\pm$246    \\
 2 &       1.00$\pm$0.05 &        19.0$\pm$0.5 &        137$\pm$2.65 &          589$\pm$32.8 &        957$\pm$52.7 &       1250$\pm$62.7 &           1450$\pm$81.8 &       3230$\pm$301 &      10400$\pm$983    \\
 3 &       2.15$\pm$0.28 &        3.84$\pm$0.6 &       6.60$\pm$0.54 &         3.95$\pm$0.40 &       2.76$\pm$0.52 &       9.58$\pm$0.75 &           42.7$\pm$3.11 &       114$\pm$11.2 &       2920$\pm$279    \\
 4 &            $-$ &            $-$ &            $-$ &         3.05$\pm$0.22 &       11.5$\pm$1.18 &       16.0$\pm$1.18 &           20.4$\pm$1.60 &       1790$\pm$167 &      31000$\pm$2900    \\
 5 &      0.18($-$) &      0.49($-$) &       1.19$\pm$0.08 &         3.62$\pm$0.23 &       9.27$\pm$0.46 &       11.1$\pm$0.54 &           12.0$\pm$0.56 &       396$\pm$36.7 &       4300$\pm$406    \\
 6 &       1.59$\pm$0.09 &       5.43$\pm$0.24 &       10.5$\pm$0.31 &         54.7$\pm$2.66 &        115$\pm$6.86 &        221$\pm$10.9 &            242$\pm$12.2 &       1160$\pm$109 &       2700$\pm$254    \\
 7 &      0.58($-$) &       1.18$\pm$0.14 &      10.1($-$) &         30.1$\pm$1.75 &       89.2$\pm$5.47 &        267$\pm$13.3 &            722$\pm$41.3 &      6860$\pm$1150 &      53800$\pm$5020    \\
 8 &      0.06($-$) &      0.45($-$) &       0.72$\pm$0.10 &         1.22$\pm$0.06 &       2.73$\pm$0.13 &       2.67$\pm$0.13 &           2.69$\pm$0.13 &       213$\pm$19.9 &       1520$\pm$150    \\
 9 &            $-$ &            $-$ &            $-$ &         0.76$\pm$0.07 &       2.26$\pm$0.13 &       2.47$\pm$0.20 &           3.03$\pm$0.15 &       191$\pm$17.7 &       1150$\pm$110    \\
10 &            $-$ &            $-$ &            $-$ &         0.12$\pm$0.01 &       0.85$\pm$0.05 &       0.83$\pm$0.05 &           0.62$\pm$0.04 &      14.7$\pm$1.37 &       3880$\pm$366    \\
11 &       0.71$\pm$0.06 &       2.69$\pm$0.15 &       5.81$\pm$0.21 &         11.1$\pm$0.66 &       17.2$\pm$0.91 &       23.3$\pm$1.13 &           32.8$\pm$1.58 &       181$\pm$16.7 &       1130$\pm$109    \\
12 &       22.4$\pm$0.39 &       88.8$\pm$1.31 &        167$\pm$2.77 &          301$\pm$16.3 &        435$\pm$26.9 &        642$\pm$31.8 &           1120$\pm$57.5 &       4040$\pm$382 &       3280$\pm$321    \\
13 &            $-$ &            $-$ &            $-$ &         0.17$\pm$0.01 &       6.40$\pm$0.32 &       36.4$\pm$1.79 &            102$\pm$5.14 &       670$\pm$63.0 &      11700$\pm$1190    \\
14 &            $-$ &            $-$ &            $-$ &         2.43$\pm$0.12 &       6.68$\pm$0.32 &       7.93$\pm$0.37 &           12.4$\pm$0.58 &       1570$\pm$147 &       5410$\pm$510    \\
15 &       95.7$\pm$2.12 &        132$\pm$3.52 &        164$\pm$2.87 &          398$\pm$23.0 &        346$\pm$21.3 &        375$\pm$18.4 &            540$\pm$26.6 &       168$\pm$15.7 &       2430$\pm$2390    \\
16 &      37.1($-$) &       36.4$\pm$1.21 &      95.2($-$) &          174$\pm$9.05 &        177$\pm$9.28 &        180$\pm$11.2 &            191$\pm$11.8 &       270$\pm$73.6 &       1670$\pm$169    \\
17 &       82.5$\pm$1.98 &        120$\pm$3.41 &        111$\pm$2.26 &         52.5$\pm$2.97 &       36.7$\pm$2.47 &       31.2$\pm$1.68 &           28.5$\pm$1.68 &      9.74$\pm$0.92 &       1040$\pm$101    \\
18 &            $-$ &            $-$ &            $-$ &         8.05$\pm$0.41 &       45.0$\pm$2.76 &       93.9$\pm$4.77 &            129$\pm$7.62 &       1320$\pm$139 &       7240$\pm$713    \\
19 &       35.5$\pm$0.78 &        123$\pm$3.62 &        246$\pm$4.75 &          481$\pm$24.5 &        786$\pm$42.6 &       1140$\pm$57.5 &            1830$\pm$138 &       4370$\pm$407 &       5190$\pm$501    \\
20 &            $-$ &            $-$ &            $-$ &         0.22$\pm$0.02 &       0.94$\pm$0.06 &       0.69$\pm$0.06 &           0.59$\pm$0.04 &      76.4$\pm$7.05 &       6590$\pm$615    \\
21 &            $-$ &            $-$ &            $-$ &         0.13$\pm$0.01 &       1.10$\pm$0.05 &       2.04$\pm$0.12 &           2.39$\pm$0.11 &      87.7$\pm$8.09 &       1750$\pm$164    \\
22 &       0.69$\pm$0.07 &       3.88$\pm$0.30 &       10.5$\pm$0.40 &         86.7$\pm$4.66 &        146$\pm$7.64 &        203$\pm$10.7 &            306$\pm$16.7 &       1140$\pm$106 &       2540$\pm$235    \\
23 &            $-$ &            $-$ &            $-$ &         0.13$\pm$0.01 &       0.42$\pm$0.04 &       0.31$\pm$0.04 &           0.16$\pm$0.03 &      4.97$\pm$0.48 &       1400$\pm$131    \\
24 &            $-$ &            $-$ &            $-$ &         0.37$\pm$0.02 &       4.58$\pm$0.24 &       11.5$\pm$0.55 &           16.3$\pm$0.79 &       295$\pm$27.3 &       9040$\pm$838    \\
25 &            $-$ &            $-$ &            $-$ &         0.25$\pm$0.03 &       1.00$\pm$0.06 &       0.98$\pm$0.07 &           0.55$\pm$0.05 &      32.4$\pm$3.04 &       2610$\pm$252    \\
26 &        263$\pm$6.29 &        344$\pm$9.82 &        305$\pm$6.19 &          177$\pm$9.40 &        141$\pm$7.04 &        140$\pm$6.81 &            213$\pm$11.2 &       590$\pm$55.2 &       1050$\pm$103    \\
27 &       0.42$\pm$0.05 &       1.67$\pm$0.09 &       4.83$\pm$0.19 &         8.36$\pm$0.44 &       9.92$\pm$0.49 &       8.98$\pm$0.43 &           7.70$\pm$0.36 &      75.9$\pm$7.03 &       1220$\pm$123    \\
28 &       1.63$\pm$0.10 &       3.36$\pm$0.18 &       4.66$\pm$0.21 &         5.10$\pm$0.30 &       4.87$\pm$0.25 &       12.6$\pm$0.62 &           42.1$\pm$2.03 &       144$\pm$13.4 &       1030$\pm$107

\label{tab2}
\enddata
\end{deluxetable}

\begin{table*}[th!]
\label{tab3}
\begin{center}
\doublerulesep2.0pt
\renewcommand\arraystretch{1.02}
\centering
\caption{AKARI FIS flux densities of YSOs selected for this study}
\begin{tabular}{ccccc}
\hline \hline
   &     N60(65 \micron) & WIDE-S(90 \micron) & WIDE-L(140 \micron) & N160(160 \micron) \\
Index & mJy & mJy & mJy & mJy \\ \hline

 1 &        3874$\pm$209 &       3638$\pm$258 &      $-\tablenotemark{(a)}$ &      $-$ \\
 2 &        12210$\pm$1140 &       12420$\pm$1360 &       14430$\pm$4470 &       21650$\pm$5660 \\
 3 &        3261$\pm$505 &       4603$\pm$104 &       5360$\pm$1720 &       $-$ \\
 4 &        80680$\pm$10600 &       133200$\pm$21000 &       124600$\pm$21200 &       217400$\pm$28800 \\
 5 &        4453$\pm$593 &       6092$\pm$420 &       12880$\pm$2490 &       10740$\pm$1110 \\
 6 &        3872$\pm$300 &       3757$\pm$320 &       6325$\pm$2010 &       10490$\pm$947 \\
 7 &        101400$\pm$1490 &       64350$\pm$11200 &       58090$\pm$10400 &       84570$\pm$885 \\
 8 &        $-$ &       2339$\pm$43 &       6789$\pm$1820 &       $-$ \\
 9 &        $-$ &       1758$\pm$162 &       7982$\pm$2490 &       $-$ \\
10 &        3121$\pm$498 &       9190$\pm$259 &       20510$\pm$3760 &       25870$\pm$3390 \\
11 &        $-$ &       2431$\pm$275 &      $-$ &       10070$\pm$182 \\
12 &        3993$\pm$333 &       4712$\pm$182 &       $-$ &       $-$ \\
13 &        15260$\pm$1370 &       25020$\pm$1530 &       46290$\pm$4150 &       57090$\pm$12500 \\
14 &        4140$\pm$252 &       4442$\pm$133 &       $-$ &       $-$ \\
15 &        6654$\pm$754 &       12640$\pm$1180 &       34980$\pm$7450 &       21750$\pm$3520 \\
16 &        2298$\pm$158 &       3590$\pm$138 &       5456$\pm$1050 &       $-$ \\
17 &        $-$ &       2574$\pm$99 &       9914$\pm$1890 &       7043$\pm$1550 \\
18 &        38380$\pm$670 &       41090$\pm$1970 &       62110$\pm$12400 &       109600$\pm$7900 \\
19 &        $-$ &       $-$ &       23600$\pm$6020 &       16030$\pm$4390 \\
20 &        5544$\pm$652 &       11470$\pm$452 &       20810$\pm$2580 &       19110$\pm$2430 \\
21 &        $-$ &       2777$\pm$130 &       9229$\pm$2700 &       $-$ \\
22 &        $-$ &       4822$\pm$632 &       14440$\pm$4450 &       8270$\pm$1480 \\
23 &        $-$ &       2164$\pm$101 &       7670$\pm$530 &       $-$ \\
24 &        8573$\pm$1530 &       15800$\pm$582 &       31160$\pm$2020 &       34310$\pm$8050 \\
25 &        $-$ &       3807$\pm$505 &       7356$\pm$938 &       $-$ \\
26 &        $-$ &       1470$\pm$236 &       $-$ &      $-$ \\
27 &        $-$ &       2146$\pm$218 &       6882$\pm$981 &       $-$ \\
28 &        $-$ &       1542$\pm$66 &      $-$ &      $-$ \\ \hline
\end{tabular}
\end{center}
\begin{tabnote}
\hskip78pt $^{\rm a}$
Fluxes of the flux density flag smaller than 2 are not considered. \\
\end{tabnote}
\end{table*}

\begin{table*}[th!]
\label{tab4}
\begin{center}
\centering
\caption{Physical properties of the selected 28 YSOs}
\doublerulesep2.0pt
\renewcommand\arraystretch{1.02}
\begin{tabular}{cccrcccc}
\hline \hline
\colhead{} & \colhead{} &  \colhead{}    &  \multicolumn{2}{c}{$T_{bol}(K)$} &   \colhead{}   &
\multicolumn{2}{c}{$L_{bol}(L_{\odot})$} \\
\cline{4-5} \cline{7-8} \\
\colhead{} & \colhead{} &  \colhead{}  &  \colhead{without} &  \colhead{with} & \colhead{}   & \colhead{without} &  \colhead{with} \\
\colhead{Index} & \colhead{Spectral Index$(\alpha)$} &  \colhead{$L_{int}(L_{\sun})$\tablenotemark{1}} &  \colhead{FIR fluxes\tablenotemark{2}}   & \colhead{FIR fluxes\tablenotemark{3}}    & \colhead{} &
\colhead{FIR fluxes\tablenotemark{2}} & \colhead{FIR fluxes\tablenotemark{3}}
\\ \hline
 1 & -0.610   &     3.371 & 1202\phn & 1143 & & 0.763 & 0.804 \\
 2 &  0.140   &    12.590 & 452\phn & 380 & & 0.919 & 1.116 \\
 3 &  0.580   &     3.814 & 180\phn & 119 & & 0.110 & 0.199 \\
 4 &  2.160   &    35.148 & 76\phn & 53 & &   3.094 & 12.593 \\
 5 &  1.090   &     5.488 & 105\phn & 75 & &  0.490 & 0.894 \\
 6 &  0.830   &     3.544 & 307\phn & 234 & & 0.846 & 1.167 \\
 7 &  2.350   &    59.015 & 119\phn & 85 & &  6.997 & 13.863 \\
 8 &  1.110   &     2.065 & 114\phn & 84 & &  0.196 & 0.282 \\
 9 &  1.510   &     1.588 & 104\phn & 67 & &  0.158 & 0.360 \\
10 &  0.980   &     4.983 & 56\phn & 41 & &   0.316 & 0.856 \\
11 &  0.400   &     1.562 & 283\phn & 145 & & 0.191 & 0.435 \\
12 &  0.410   &     4.255 & 496\phn & 458 & & 2.822 & 3.078 \\
13 &  3.330   &    14.064 & 92\phn & 61 & &   1.211 & 2.973 \\
14 &  1.930   &     6.811 & 114\phn & 107 & & 0.962 & 1.066 \\
15 & -0.980   &     3.209 & 1015\phn & 549 & &1.300 & 2.489 \\
16 & -0.850   &     2.256 & 847\phn & 641 & & 0.730 & 0.981 \\
17 & -1.920   &     1.445 & 1559\phn & 1033 & & 0.466 & 0.711 \\
18 &  1.700   &     8.957 & 138\phn & 64 & & 1.199 & 6.191 \\
19 &  0.260   &     6.550 & 546\phn & 448 & & 4.168 & 5.168 \\
20 &  1.660   &     8.199 & 59\phn & 47 & & 0.798 & 1.640 \\
21 &  1.910   &     2.357 & 76\phn & 50 & & 0.246 & 0.502 \\
22 &  0.670   &     3.346 & 334\phn & 240 & & 0.710 & 1.037 \\
23 &  0.700   &     1.911 & 56\phn & 36 & & 0.164 & 0.311 \\
24 &  2.150   &    11.036 & 72\phn & 53 & & 1.200 & 2.552 \\
25 &  1.100   &     3.432 & 60\phn & 47 & & 0.141 & 0.227 \\
26 & -0.620   &     1.458 & 1492\phn & 1430 & & 0.491 & 0.513 \\
27 & -0.190   &     1.679 & 232\phn & 136 & & 0.051 & 0.098 \\
28 &  0.920   &     1.432 & 276\phn & 323 & & 0.059 & 0.048 \\ \hline
\end{tabular}
\end{center}
\begin{tabnote}
\hskip28pt $^{\rm 1}$
{The intenal luminosity calculated from the flux at 70 $\micron$ (Dunham et al. 2008).} \\
\end{tabnote}
\vspace{0.1cm}
\begin{tabnote}
\hskip28pt $^{\rm 2}$
$T_{bol}$ and $L_{bol}$ calculated only with the c2d data. \\
\end{tabnote}
\vspace{0.1cm}
\begin{tabnote}
\hskip28pt $^{\rm 3}$
{$T_{bol}$ and $L_{bol}$ calculated with the c2d data combined with the AKARI/FIS data.} \\
\end{tabnote}
\end{table*}

\section{DATA CRITERIA}

The sample YSOs selected for this study satisfy three criteria: 1. They are listed in the c2d catalogs, 2. Their MIPS2 70 $\micron$ fluxes are greater than 1 Jy, and
3. They are detected by the AKARI but not by the IRAS.

We use YSOs from the c2d YSO catalogs as our study targets.
Considering the sensitivity limit of the AKARI/FIS, only those YSOs with 70 $\micron$ fluxes greater than 1 Jy are used.
The Key catalog for this study is the AKARI/FIS Bright Source Catalog since the goal of our study is to understand how FIR fluxes affect the properties of YSOs derived from observations.
The IRAS catalog could be used to study the effect of the FIR, but we adopt the AKARI/FIS catalog
since the AKARI resolved the point sources effectively due to its higher resolution (Jeong et al. 2007) and absolute calibration errors are expected to be less than 20\% in all bands (Yamamura et al. 2010). In addition, most of 100 $\micron$ fluxes for YSOs matched with IRAS  do not have a good flux quality. Therefore, we have used only AKARI/FIS sources with a good flux quality for our study.
Besides, previous studies of those YSOs detected in AKARI but not detected in IRAS have been conducted without the FIR data.
Therefore, it is meaningful to compare previous studies with our study, which includes FIR fluxes.

We carry out cross correlation with 6$\arcsec$ among catalogs based on the coordinates provided in the c2d YSOs catalogs considering positional uncertainties (Yamamura et al. 2010).
There is little flux dependency of the position error on the current flux range of the sample.
The AKARI/FIS catalog provides the flux density quality flag, which indicates whether observed flux density is reliable or not.
For our study, we use only the fluxes flagged as ``reliable'' from ``confirmed'' sources (Yamamura et al. 2010).
Based on these criteria, 28 YSOs have been selected as our study sample, as listed in Table~\ref{tab1}.
The flux densities and their errors of these 28 YSOs are presented in Table~2 and 3.

\section{EVOLUTIONARY INDICATORS}

In order to classify the evolutionary stages of YSOs, a few evolutionary indicators have been used. We summarize them below.

\subsection{Spectral Index}

Since Lada \& Wilking (1984) recognized that the SEDs of YSOs were falling into natural groups, Lada (1987) first codified
the tripartite class system using the spectral index ($\alpha$) between 2 and 20 \micron:
   \begin{equation}\alpha=\frac{d\rm{log}(\lambda S_{\lambda})}{d\rm{log}(\lambda)}\end{equation}
where $\lambda$ is the wavelength and $S_{\lambda}$ is the flux density at that wavelength.
The boundaries of class system by Lada (1987) are as follows:

 \textbf{I} \ \ 0  $<$ $\alpha$ $\lesssim$ 3,

 \textbf{II} \ \ -2 $<$ $\alpha$ $\lesssim$ 0, and

 \textbf{III} \ \ -3 $<$ $\alpha$ $\lesssim$ -2.

After the tripartite class system by Lada (1987) was introduced, the class system has been improved by Wilking et al. (1989),
Andr\'{e} et al. (1993, 2000) and Greene et al. (1994) who formalized the 4-class system, as follows:

  \textbf{I} \ \ 0.3 $\lesssim$ $\alpha$,

  \textbf{Flat} \ \ -0.3 $\lesssim$ $\alpha$ $<$ 0.3,

  \textbf{II} \ \ -1.6  $\lesssim$ $\alpha$ $<$ -0.3, and

  \textbf{III} \ \ $\alpha$ $<$ -1.6.

Although this class system was based on the slope between 2.2 and 10 $\micron$,
they asserted that these classes showed no systematic deviations from classes based on the slopes between 2.2 and 20 $\micron$.
The c2d catalogs provide a least-squares fit to all photometry between 2 and 24 $\micron$ to determine
the spectral index ($\alpha$) and classify objects into the four classes defined by Greene et al. (1994) (Evans et al. 2009).
We use $\alpha$ provided by the c2d catalogs as an evolutionary indicator of our selected YSOs.

\begin{figure}[!!t]
\epsscale{.98}
\plotone{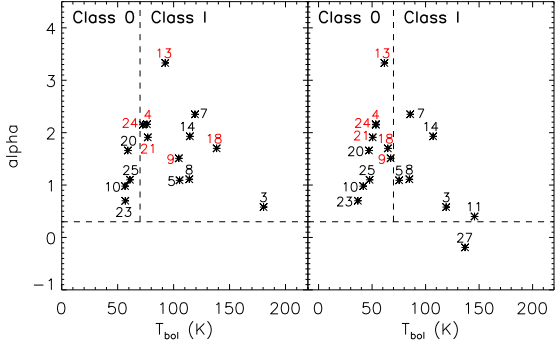}
\caption{The plot of $\alpha$ vs. $T_{bol}$ for a subset of the sources from Table~4. We present sources only with $T_{bol}$ $<$ 200 K in this figure.
The numbers in this figure indicate the index numbers of YSOs designated in Table~1, 2, and 3.
The horizontal dashed line indicates the boundary between the embedded (class 0/I) and less embedded (class II) sources,
and the vertical dashed line indicates the boundary between class 0 ($T_{bol}$ $<$ 70 K) and class I (70 K $<$ $T_{bol}$ $<$ 650 K).
The left panel shows the values of $T_{bol}$ calculated without the AKARI FIR data,
and the right panel shows those calculated with the AKARI FIR data as well.
The numbers of red color denote the YSOs whose evolutionary stage changed from class I to class 0 in this comparison. \label{fig1}}
\end{figure}

\subsection{Bolometric Temperature}

Myers $\&$ Ladd (1993) suggests $T_{bol}$ as a measure of the circumstellar obscuration and evolutionary development
of a YSO. We calculate the bolometric temperature $(T_{bol})$, which is the temperature of a blackbody having the same mean
frequency $\bar{\nu}$ as the observed continuum spectrum, using the prescription given in Myers $\&$ Ladd (1993),
 \begin{equation} T_{bol}\equiv [\zeta(4)/4\zeta(5)] h \bar{\nu} /k = 1.25 \times 10^{-11} \bar{\nu} \, K Hz^{-1},
\end{equation}

where $\zeta$ (n) is the Riemann zeta function of argument n, h is Planck's constant, and k is Boltzmann's constant.
The mean frequency $\bar{\nu}$ is the ratio of the first and zeroth frequency moments of the spectrum,
 \begin{equation} \bar{\nu} \equiv \frac{\int^{\infty}_{0} \nu S_{\nu}d\nu}{\int^{\infty}_{0} S_{\nu}d\nu}.
\end{equation}

Chen et al. (1995) showed that evolutionary classes, including class 0, ranged depending on $T_{bol}$ as follows:

 \textbf{0} \,\,$T_{bol}$ $<$ 70,

 \textbf{I} \,\,70 $\lesssim$ $T_{bol}$ $\lesssim$ 650, and

 \textbf{II} \,\, 650  $<$ $T_{bol}$ $\lesssim$ 2800.

\subsection{The ratio between $L_{bol}$ and $L_{smm}$}

The ratio of bolometric to submillimeter luminosity ($L_{bol}/L_{smm}$) can be an evolutionary indicator of YSOs, especially in the early embedded phase (Andr\'{e} et al. 1993).
$L_{bol}$ is calculated by integrating over the full observed SED,
 \begin{equation} L_{bol}=\int ^{\infty} _{0} 4 \pi D^{2} S_{\nu} d \nu, \end{equation}
while the submillimeter luminosity is calculated by integrating over the observed SED for $\lambda \geq 350 \ \ \micron$,
 \begin{equation} L_{smm}=\int ^{\nu=c/350 \micron} _{0} 4 \pi D^{2} S_{\nu} d \nu. \end{equation}
where $S_{\nu}$ is the flux density and D is the distance. Young \& Evans (2005) suggested these boundaries for classifying protostars:
$L_{bol}/L_{smm}$ = 35 for Pre-Protostellar Core/Class 0, $L_{bol}/L_{smm}$ = 175 for Class 0/I, and $L_{bol}/L_{smm}$ $\sim$ 2000 for the Class I/II transition.

The spectral index $\alpha$ is calculated without the FIR data so that it cannot be used to subdivide the embedded stage into Class 0 and Class I.
However, $T_{bol}$ and $L_{bol}$/$L_{smm}$ have ability to distinguish Class 0 from Class I.
A drawback to using $L_{bol}$/$L_{smm}$ is in the lack of the complete data set at $\lambda$ $\geq$ 350 $\micron$ in many YSOs.

\section{RESULTS}

We calculated $T_{bol}$ and $L_{bol}$ of 28 YSOs selected by the criteria presented in section 3
in both cases of (1) only with NIR and MIR and (2) including the AKARI-FIR data as well.
Table~4 lists the spectral index provided from the c2d catalogs and $T_{bol}$ and $L_{bol}$ calculated by integrating the observed flux density within the wavelengths
covered by observations.

We calculated the internal luminosity ($L_{int}$) listed in Table~4 using the equation presented in Dunham et al. (2008):
\begin{equation} L_{int}=3.3 \times 10^{8}F^{0.94}_{70} L_{\sun}, \end{equation}\\
where $F_{70}$ is a flux density at 70 $\micron$ in cgs units (ergs $cm^{-2}$ $s^{-1}$) and normalized to 140 pc. This equation is derived using the results of the linear least-square fit to the modeled $L_{int}$ versus
$F_{70}$ in log-log space. Since $F_{70}$ is an observable quantity, the correlation they found between $F_{70}$ and $L_{int}$ can be used to estimate $L_{int}$ for sources which either lack
sufficient data to constrain radiative transfer models or have such data but have not yet been modeled.

\begin{figure}[!!!!t]
\epsscale{.98}
\plotone{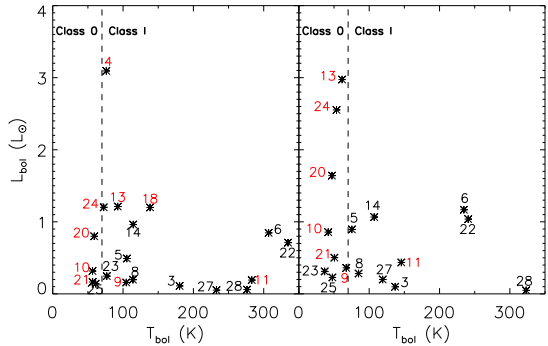}
\caption{The plot of $L_{bol}$ vs. $T_{bol}$ for the sources from Table~4. The numbers denote the index of YSOs as listed in Table~\ref{tab1}.
The categories of two panels are the same as in Fig.~\ref{fig1}, that is, without (left panel) and with (right panel) the FIR fluxes in the calculation
of $L_{bol}$. The dashed line indicates the boundary between class 0 and class I in $T_{bol}$. The numbers of red color denote the YSOs whose
$L_{bol}$ increases more than twice when the FIR fluxes are included in the calculation of $L_{bol}$. \label{fig2}}
\end{figure}

We compare the values of $T_{bol}$ of the 28 YSOs calculated with and without the AKARI FIR fluxes in Fig.~\ref{fig1}.
According to $\alpha$, 20 out of 28 YSOs are classified as Class 0/I.
As mentioned, in the previous section, however, $\alpha$ cannot subdivide the embedded phase. Therefore, for the detail study of the early embedded YSOs,
we need to use $T_{bol}$ and $L_{bol}$/$L_{smm}$. As presented in Fig.~\ref{fig1}, class 0 and class I objects can be divided based on
$T_{bol}$; $T_{bol}$ $<$ 70 K for class 0 and $T_{bol}$ $>$ 70 K for class I. When the FIR fluxes are not included in the calculation of $T_{bol}$ (the left
panel of Fig.~\ref{fig1}, see Table~4 for the whole samples), 5, 19, and 4 YSOs are classified to class II, class I, and class 0, respectively. However, when we
include the FIR fluxes in the calculation of $T_{bol}$ (the right panel of Fig.~\ref{fig1}), 3, 15, and 10 YSOs are classified to class II, I, and 0, respectively.

That is, 6 YSOs are reclassified to class 0 from class I, and 2 YSOs are reclassified to class I from class II. As a result, YSOs detected in the FIR wavelengths are
mostly (25 out of 28) embedded. This comparison shows that the FIR fluxes are very important to know the evolutionary stages of YSOs, especially for the
embedded ones.

\begin{figure}[!!!!!t]
\epsscale{.98}
\plotone{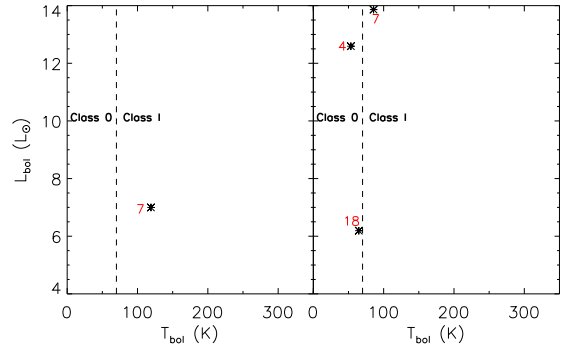}
\caption{The same plot as Fig.~\ref{fig2} but for YSOs with higher $L_{bol}$. The values of $L_{bol}$ of sources 4 and 18 increase more than
4 times when the FIR fluxes are included in the calculation of $L_{bol}$. \label{fig3}}
\end{figure}

Fig.~\ref{fig2} and~\ref{fig3} present the same comparison but for $L_{bol}$. When the FIR fluxes are included in the calculation of $L_{bol}$, the values of $L_{bol}$
of 10 YSOs increase by more than a factor of 2. Two of them (sources 4 and 18) even increase by more than a factor of 4 in their values of $L_{bol}$.
This change in $L_{bol}$ possibly increases the ratio, $L_{bol}$/$L_{smm}$, resulting in different evolutionary stages of the YSOs.
According to our comparisons in the calculations of $T_{bol}$ and $L_{bol}$ with and without the FIR fluxes, we can see how important knowing the FIR fluxes of YSOs,
especially the embedded ones, is.

\section{ONE-DIMENSIONAL DUST RADIATIVE TRANSFER MODELING}
The original radiation from the central source and circumstellar disk is scattered, absorbed and reemitted by the dusty material in the envelope.
The embedded sources have thick envelopes, where most of mass resides, and which can be well described by a sphere.
Therefore, in order to get the detailed information on YSOs, the emerged SED must be modeled. In this study, we modeled the SED of a YSO using a 1-D radiative transfer code
to test how the FIR fluxes can constrain the physical parameters of the YSO.

The selected YSO for this test is the source 13 in Table~\ref{tab1}. Its Spitzer source name is J033317.9+310932, and it is located in a large molecular cloud, Perseus.
As listed in Table~4, $L_{bol}$ is $\sim$1.2 and $\sim$3.0 $L_{\sun}$ when calculated without and with the FIR fluxes, respectively.
In the early embedded phase, the heating by the ISRF cannot be ignored compared to the heating by the central source (Young et al. 2003). Therefore, in order to know the accurate internal luminosity, which is a crucial quantity to
understand the formation mechanism of the protostar, separately from the ISRF luminosity, we must model the observed SED.

J033317.9+310932 was not detected by the 2MASS observations.
This source is listed as Bolo 80 in the Bolocam survey of Perseus for 1.1 mm dust continuum emission
 (Enoch et al. 2006).
The observed fluxes and calculated $T_{bol}$ and $L_{bol}$ of this YSO are listed in Table~2, 3, and 4.
We use the one-dimensional radiative transfer code, DUSTY (Ivezic et al. 1999) to model the SED of J033317.9+310932.
The program, ObsSphere (Shirley et al. 2002) is then used to match the results from DUSTY with the actual observed fluxes at wavelengths longward of 100 $\micron$.
ObsSphere calculates the fluxes emerging only from the envelope, which is the dominant flux source at long wavelengths, based on the physical conditions derived from
DUSTY. Refer to Dunham et al. (2006) for the detailed description of the model input parameters.

We modeled the SED with and without the FIR fluxes and compared the derived properties of the central heating source (star/disk) and the envelope.
The envelope is assumed to have a power-law radial density profile $[n(r)\infty r^{-p}]$.
We have tested the envelope model by changing the p parameter from 1.5 to 2.0.
According to the inside-out collapse model (Shu et al. 1977),
the density profile has the power of 1.5 and 2 at radii smaller and larger than the infall radius, respectively.
The heating sources for our model are the internal star/disk system and the ISRF. In general, since cores with YSOs are embedded in bigger clouds,
the ISRF reaching the surface of a core is attenuated by the surrounding material.
We tested various attenuator of the ISRF, which is parameterized by the visual extinction ($A_V$) since the UV field strength of the ISRF relative to the local ISRF,
$G_{0}$ is given by $e^{-1.8A_V}$ (Lee et al. 2004).

\begin{figure}[!!t]
\epsscale{.98}
\plotone{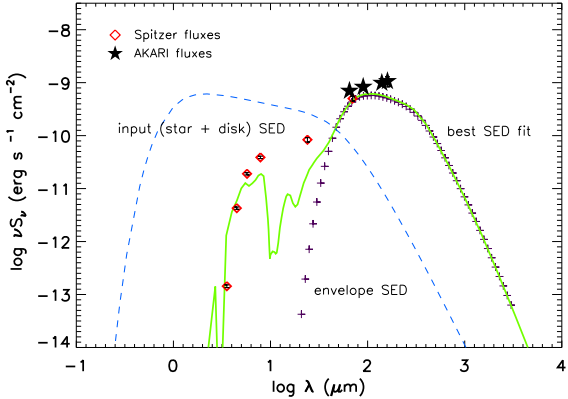}
\caption{The best-fit model of the SED of J033317.9+310932 when only the Spitzer data are used.
The green solid line indicates the best-fit model SED obtained by DUSTY, and red open diamonds represent the Spitzer photometry data.
The blue dashed line shows the input SED for the model.
The purple crosses show the emerging SED only from the envelope which is calculated with ObsSphere based on the model density profile and the temperature profile obtained from DUSTY. The star symbols denote the AKARI fluxes, which are not included for this SED model.
The best-fit model only with the Spitzer fluxes does not match well the AKARI fluxes. \label{fig4}}
\end{figure}

For the dust opacity, as many studies did (Shirley et al. 2002; Young et al. 2003; Evans et al. 2005), we adopted the OH5 dust model,
which is for the thin ice mantles coagulated for $10^{5}$ years at a gas density of $10^{6}$ $cm^{-3}$ (Ossenkopf $\&$ Henning 1994).

To find the best-fit envelope model we calculated the reduced $\chi^{2}$ (Dunham et al. 2006), 

 \begin{equation} \chi^{2}=\frac{1}{n-1} \displaystyle\sum\limits_{i=0}^n \frac{[S^{obs}_{\nu}(\lambda_{i})-S^{mod}_{\nu}(\lambda_{i})]^{2}}{\sigma_{\nu}(\lambda_{i})^2}, \end{equation} \\

where $S^{obs}_{\nu}(\lambda_{i})$ is the observed flux density at $\lambda_{i}$, $S^{mod}_{\nu}(\lambda_{i})$ is the simulated flux density from this model at same wavelength,
and $\sigma_{\nu}(\lambda_{i})$ is the uncertainty in the observed flux density.
And n is the number of the data points, so the model with the FIR fluxes has four more data points than the other model without the FIR fluxes for the case of J033317.9+310932.

Fig.~\ref{fig4} shows the best-fit model without the FIR fluxes. The model parameters of this model are listed in Table~\ref{tab5}.
The best-fit model does not match very well the observed 8 and 24 $\micron$ fluxes. This is possibly because the disk emission is not well modeled with our 1-D calculation.
As seen in Fig.~\ref{fig4}, this best-fit SED cannot match the AKARI fluxes.
According to this best-fit model, the ISRF is completely obscured by surrounding medium, and the internal luminosity is about 3.8 $L_{\sun}$.

\begin{figure}[!!t]
\epsscale{.98}
\plotone{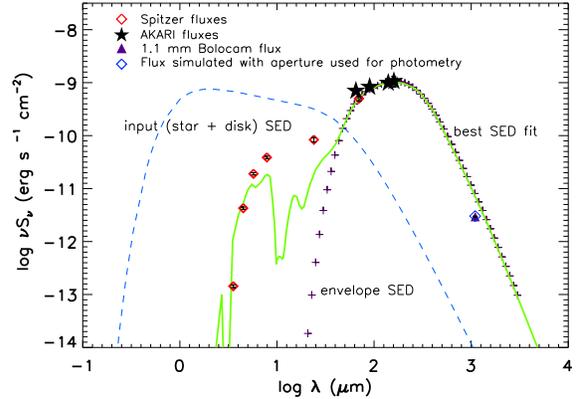}
\caption{The best-fit model SED when the AKARI FIR fluxes as well as the Spitzer data are included.
The purple triangle and open blue diamond indicate the observed 1.1 mm Bolocam flux and the flux simulated from the model with the aperture used for the 1.1 mm flux measurement, respectively. Other colors and symbols are the same as in Fig.~\ref{fig4}. \label{fig5}}
\end{figure}

Fig.~\ref{fig5} presents the best-fit model when the FIR fluxes are included. The parameters of this model are also listed in Table~\ref{tab5}.
The internal luminosity increases $\sim$25$\%$ compared to the model without the FIR fluxes, and a higher UV field strength of the ISRF is necessary to fit the AKARI fluxes.
The model SED shows that the peak of the SED is located at the wavelength of $\sim$160 $\micron$, which is the longest wavelength of the AKARI FIS.
In order to constrain the SED better, however, observations at longer wavelengths than the AKARI bands are necessary.

\begin{deluxetable}{cccccccccc}
\small
\tablecolumns{10}
\tablewidth{0pt}
\tablecaption{The best-fit models for the SED of J033317.9+310932}
\tablehead{
\colhead{model} & \colhead{p\tablenotemark{1}} & \colhead{$R_{inner}$\tablenotemark{2}} & \colhead{$R_{outer}$\tablenotemark{3}} &
\colhead{$M_{env}$\tablenotemark{4}} & \colhead{$n_{1000 AU}$\tablenotemark{5}} & \colhead{$G_{0}$\tablenotemark{6}} & \colhead{$L_{int}$\tablenotemark{7}} & \colhead{$T_{star}$\tablenotemark{8}} \\
\colhead{} & \colhead{$-$}  & \colhead{AU} & \colhead{AU} & \colhead{$M_{\sun}$}   & \colhead{$cm^{-3}$}  &
\colhead{$-$} & \colhead{$L_{\sun}$}  & \colhead{K}}
\startdata

without FIR fluxes & 1.6 & 650 & 20000 & 5.9 & 1.5$\times10^{6}$ & 0 & 3.8 & 2800 \\
with FIR fluxes & 1.9 & 880 & 29000 & 7.6 & 2.4$\times10^{6}$  & 0.4 & 4.7 & 3000

\label{tab5}
\enddata
\tablenotetext{1}{The power of the power law envelope density profile.}
\tablenotetext{2}{The inner radius of the envelope.}
\tablenotetext{3}{The outer radius of the envelope.}
\tablenotetext{4}{The mass of the envelope.}
\tablenotetext{5}{The fiducial density at 1000 AU.}
\tablenotetext{6}{The UV field strength of the ISRF relative to the local ISRF.}
\tablenotetext{7}{The internal luminosity caused by the accretion to the star/disk system.}
\tablenotetext{8}{The temperature of the central protostar.}
\end{deluxetable}

\begin{figure}[!!t]
\epsscale{.98}
\plotone{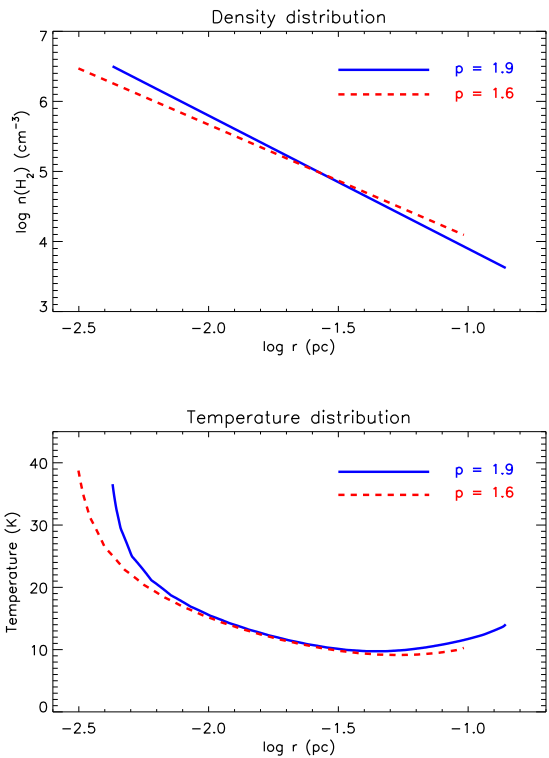}
\caption{The envelope physical structures obtained from the dust continuum modeling as a function of radius. The upper and lower panels present
the density and temperature structures, respectively. The red dashed lines are for the best-fit model without the FIR fluxes
while the blue solid lines are for the best-fit model with the FIR fluxes. \label{fig6}}
\end{figure}

The density (upper panel) and temperature (lower panel) distributions of the envelope with radius
for the best-fit models with (blue solid line) and without (red dashed line) the FIR fluxes
are presented in Fig.~\ref{fig6}.
Since AKARI FIR fluxes listed in the catalogue are not color corrected ones, we applied color correction to AKARI FIR fluxes by using our modeled SEDs.
The variation of color correction factor in our models does not exceed 10$\%$.
The envelope density profile of the best-fit model with the FIR fluxes has the power of 1.9, which is similar to the density profile of the static envelope
in the Shu's inside-out collapse model. This result might indicate that J033317.9+310932 is in a very early stage of collapse, so the infall radius is still small.

The physical parameters for the best-fit model with the FIR fluxes show differences from those for the best-fit model without the FIR fluxes.
As mentioned above, the power of the density profile is different in the two models. In addition, the envelope mass of the best-fit model with the FIR fluxes is
about two solar mass greater than the best-fit model without the FIR fluxes.
Other important parameters are the internal luminosity ($L_{int}$) and the UV strength of the ISRF ($G_{0}$), which are two
heating sources of the envelope. According to our test, $L_{int}$ and $G_{0}$, constrained by the SED modeling, are probably close to $L_{int}$ = 4.7 $L_{\sun}$ and $G_{0}$ = 0.4
rather than $L_{int}$ = 3.8 $L_{\sun}$ and $G_{0}$ = 0. Here, $G_{0}$ = 0.4 is equivalent with the UV field attenuated by $A_V$ = 0.5 mag.

\section{SUMMARY AND CONCLUSIONS}

We tested the effect of FIR fluxes in classification of the evolutionary stages of YSOs and in calculation of the internal luminosity as well as
the UV strength of the external heating source, the ISRF. Twenty eight sample YSOs, which have their counterparts only in the AKARI/FIS
Bright Source Catalog, were selected from the c2d YSO catalogs.
First, we calculated an evolutionary indicator, $T_{bol}$, and found that 6 objects, which are classified into class I when only Spitzer fluxes were used,
were reclassified into class 0 when the AKARI FIR fluxes were included.
Two objects, which were classified into class II when only the Spitzer data were included, were also reclassified into class I.
This result shows that the FIR observations are very important in the classification of the evolutionary stages of the early embedded YSOs since they emit most of their energy
at the long wavelength regime.

In order to test how the FIR fluxes constrain the physical structure of the envelope, the internal luminosity, and the UV strength of the ISRF,
we modeled the SED of one of our sources.
In the early evolutionary phase, the external heating by the ISRF can not be ignored compared to the internal heating.

According to our dust continuum modeling, when the FIR fluxes are included, the density profile of the envelope is steeper (p=1.9 compared to p=1.6)
indicating most of envelope material is possibly static, that is, the infall radius is very small in the sense of Shu's inside-out collapse model.
The estimated internal luminosity is constrained to 4.7 $L_{\sun}$ when the FIR fluxes are included (compared to 3.8 $L_{\sun}$ obtained when the FIR fluxes are not included).
The UV strength of the ISRF is $G_{0}$=0.4 (attenuated by $A_V$ = 0.5 mag) in the model with the FIR fluxes, but in the model without the FIR fluxes,
the ISRF needs to be completely attenuated ($G_{0}$=0).
Our modeling result also indicates that fluxes at FIR and even longer wavelengths are crucial to better constrain the physical parameters of
embedded YSOs.

\vspace{0.5cm}

\acknowledgments
This research was supported by the National Research Foundation of Korea(NRF) grant funded by the Korea government(MEST) (No. 2009-0062866) and by Basic Science Research Program through the NRF funded by the Ministry of Education, Science and Technology (No. 2010-0008704). This research is based on observations with AKARI, a JAXA project with the participation of ESA.

\end{document}